\documentclass[prl,aps,preprint,groupedaddress,floats,showpacs]{revtex4}

\usepackage{graphicx}

\usepackage{amssymb}

\begin{document}





\title{Deconfined criticality, runaway flow in the two-component
scalar electrodynamics and weak first-order superfluid-solid transitions}


\author{A. B.
Kuklov${^1}$,
N. V.
Prokof'ev$^{2,3,4}$, B. V.
Svistunov$^{2,4}$,
and M.
Troyer$^5$}

\address{${^1}$Department of Engineering Science and
Physics, The College of Staten Island, CUNY, Staten
Island, NY 10314\\
${^2}$Department of Physics, University of
Massachusetts, Amherst, MA 01003 \\
$^3$Dipartimento di Fisica, Universit\`a di Trento and CNR-INFM BEC Center,
I-38050 Povo, Trento, Italy\\
${^4}$Russian Research Center ``Kurchatov Institute'', 123182,
Moscow\\
${^5}$ Theoretische Physik, ETH Z\"urich, CH-8093 Z\"urich, Switzerland}

\begin{abstract}
We perform a comparative Monte Carlo study of the easy-plane
deconfined critical point (DCP) action and its short-range
counterpart to reveal close similarities between the two models
for intermediate and strong coupling regimes.
For weak coupling, the structure of the phase diagram depends 
on the interaction range: while the short-range model
features a tricritical point and a continuous U(1)$\times $U(1) transition,
the long-range DCP action is characterized by the runaway
renormalization flow of coupling into a first (I) order phase transition.
We develop a ``numerical flowgram" method for high precision
studies of the runaway effect, weakly I-order transitions, and
polycritical points. We prove that the easy-plane DCP action
is the field theory of a weakly I-order phase transition between the
valence bond solid and the easy-plane antiferromagnet (or superfluid, in
particle language) for any value of the weak coupling strength. Our analysis
also solves the long standing problem of what is the
ultimate fate of the runaway flow to strong coupling in the theory
of scalar electrodynamics in three dimensions with U(1)$\times$U(1)
symmetry of quartic interactions.
\end{abstract}


\maketitle

\section{Introduction}
\label{Intro}

Critical properties of systems described by several complex fields
coupled to the gauge field have a long history of studies and are
relevant for numerous problems in physics which include
normal-superfluid transitions in multi-component neutral or
charged liquids (see, e.g. \cite{Halperin2,Brezin,Chen,Babaev}),
superfluid--valence-bond solid (SF-VBS) transitions in lattice
models \cite{dcp1,dcp2,dcp3}, the Higgs mechanism in particle
physics \cite{Higgs}, etc. Recently, the authors of
Refs.~\cite{dcp1,dcp2,Motrunich} argued that the SF--VBS
transition in a $(2+1)$-dimensional system is an example of a
qualitatively new type of quantum criticality (``decondined
criticality'') that does not fit the Ginzburg-Landau-Wilson (GLW)
paradigm. As suggested in reference \cite{Zoller_Fisher}, an
experimental realization of the SF-VBS transition is possible in
the system of ultracold atoms trapped in optical lattice. If true,
the significance of the claim that DCP opens a new era in the
theory of phase transitions is hard to overestimate. The study of
the SF-VBS transition within the DCP theory framework, and the
closely related problem of the runaway flow to strong coupling in
the theory of scalar quantum electrodynamics in three dimensions
(3D) \cite{Halperin2,Brezin,Chen} is the main focus of this work.

We also discuss a short-range analog of the DCP action, namely,
the model describing lattice phases of the two-component
Bose-Hubbard (BH2) system with large integer filling factors
\cite{twocolor}. The only difference between the DCP and BH2
models lies in the interaction range. The DCP action is
characterized by a long-range  Coulomb type interaction, which is
replaced with the contact $\delta$-functional form in the BH2
analog. Despite this difference, the gross features of the phase
diagram are surprisingly similar in both models, especially in the
regimes of intermediate and strong coupling between the two field
components. Even quantitatively, the phase diagrams are nearly
indistinguishable, see Fig.~\ref{fig1}. Both feature a bicritical
point (BP) above which there are three phases: a superfluid (SF),
a paired phase and an insulator. The superfluid is characterized
by the order parameters for two complex scalar fields, 
$\langle \psi_1 \rangle \neq 0,\, \langle \psi_2 \rangle \neq 0$. 
In the paired phase the order parameter is 
$\langle \psi^*_1 \psi_2 \rangle \neq
0$ while $\langle \psi_1 \rangle = \langle \psi_2 \rangle = 0$. In
the context of the DCP theory, the paired phase is associated with
the valence-bond supersolid (SFS); within the BH2 model
\cite{twocolor} it corresponds to the super-counter-fluid (SCF)
phase describing pairing between particles of one component and
holes of another \cite{SCF}. Finally, the insulating phase
represents the VBS state in the DCP theory and the Mott insulator
(MI) state in the BH2 model. In this phase all off-diagonal order
parameters are zero.

The crucial difference between the two models is seen {\it below}
the bicritical point BP, where direct transitions from insulator
to  SF take place. In the DCP action this line describes the
SF-VBS transformation. An implicit assumption made in
Refs.~\cite{dcp1,dcp2,Motrunich} was that as the interaction
constant between the field components, $g$, weakens down to $g=0$
(where DCP and BH2 actions become identical and describe two
decoupled XY-models), there exists a lower tricritical point (TP)
$g_{LTP}>0$ below which the VBS-SF transition becomes continuous.
Since lower TP does exist in the short-range BH2 model
\cite{twocolor}, the expectation was that both models have
essentially identical phase diagrams, with only one important
difference--- in the short-range case the continuous transition below
lower TP accounts for the U(1)$\times$U(1) universality, while the
continuous transition in the DCP action for $0< g< g_{LTP}$ would
be in the new ``deconfined" universality class. However, as we
suggested earlier \cite{YKIS} and now clearly demonstrate below,
no such  lower TP exists in the DCP action and the line of I-order
SF-VBS transitions extends all the way to $g\to 0$.

The tool we introduce for studying subtle features of the phase
diagram is the numerical flowgram method. It is based on scaling
properties of quantities similar to Binder cummulants
\cite{Binder} and turns out to be much more adequate for analyzing
weakly I-order transitions than the conventional method based on
the bimodal shape of the energy/action distribution and hysteresis
loops. The flowgram technique can also be used for systematic
identification of polycritical points. To the best of our
knowledge no other method can achieve this goal with comparable 
accuracy. First, we
construct the flowgram for the BH2 model and demonstrate its
efficiency for those parts of the phase diagram which can be
reliably identified by more conventional methods. Next, we apply
the new method to the most controversial region of the BH2 phase
diagram between BP and upper TP, which we failed to
resolve in the previous study \cite{twocolor}. The existence of
upper TP proves that the mean field does reproduce the topology of
the phase diagram for the short-range model correctly, with the
reservation that fluctuation-induced effects significantly shorten
the interval between the bicritical and the upper tricritical
points.

Finally, we apply the flowgram method to the self-dual version of
the DCP action \cite{Motrunich}. We have identified the bicritical
and the upper tricritical points, but no lower TP has been found.
We directly observe the runaway flow to strong coupling and prove
that its ultimate fate is a I-order transition. More specifically,
for small $g$, we observe data collapse with respect to the
coupling strength rescaled by the system size $L$. This collapse
indicates the existence of an effective long-range coupling
$g_{\rm eff}(L)$ and proves that it is never weak in the $L\to
\infty$ limit: no matter how small is  $g$ at short scales it
always gets renormalized to $ g_{\rm eff} \sim 1$ where the I-order 
transition takes place. Perfect data collapse continuously
relating small $g$ to one for which I-order transition is observed 
at accessible sizes provides crucial
evidence against lower TP and in support of the claim that the
direct VBS-SF transition is always weakly I-order in accordance
with the Ginzburg-Landau-Wilson paradigm \cite{YKIS}.

On one hand, our results represent the first solution of the
long-standing runaway problem for the two-component field-theory
in question. On the other hand, they explain the outcome of
recent numerical simulations which report either difficulties
with finite-size scaling of the data \cite{Sandvik} or weakly I-order
transitions \cite{weak1} in microscopic models of the SF-VBS
transition.

\section{DCP action and its short-range counterpart}

\begin{figure}
\begin{center}
\includegraphics[angle=-90, width=0.8\columnwidth]{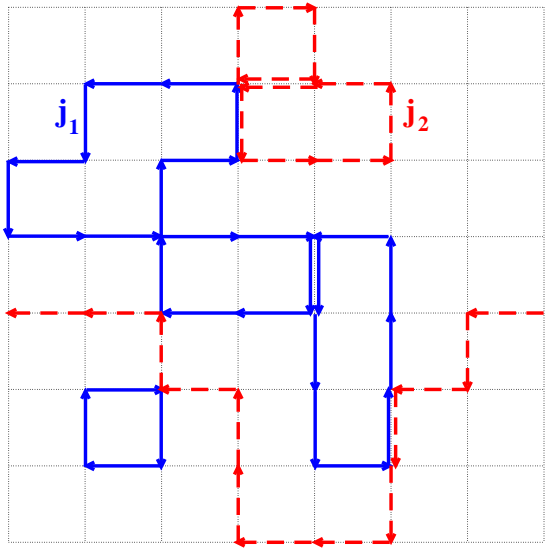}
\caption{(Color online). A typical configuration of currents
for the two-component $j$-current model. The configuration space
consists of arbitrary number of closed oriented loops which are allowed
to overlap and intersect. The current ${\bf j}_{a=1,2}$ is defined
as an algebraic sum of all same-type arrows on the corresponding bond.
} \label{fig0}
\end{center}
\vspace*{0.5cm}
\end{figure}

There are many equivalent formulations of the action describing
multi-component scalar fields coupled by the Abelian gauge field
\cite{dcp1,Babaev,Motrunich}. In what follows we will concentrate on
the case of two identical non-convertible components and will 
employ the integer-current lattice
representation which can be viewed as either a high-temperature
expansion for XY models in three dimensions, or as a path-integral
(world-line) representation of the interacting quantum system in
discrete imaginary time in $(d+1)=3$ dimensions. The DCP action
is explicitly symmetric with respect
to exchanging the components. After the gauge field is integrated out,
it reads
\begin{equation}
S_2^{(l)} =U {\bf j}_{1{\bf r}}^2 + U {\bf j}_{2{\bf r}}^2 + g
Q_{{\bf r}- {\bf r}'}
 \left( {\bf j}_{1{\bf r}}+{\bf j}_{2{\bf r}} \right) \cdot
 \left( {\bf j}_{1{\bf r}'}+{\bf j}_{2{\bf r}'} \right) \;,
\label{DCPA}
\end{equation}
with $U>0,\, g>0$.
The lattice Fourier transform of the long-range interaction
potential $Q_{{\bf r}- {\bf r}'}$ is given by  $Q_{\bf q} =
1/\sum_{\mu } \sin^2 (q_{\mu}/2)$, which implies the asymptotic
form $Q(r \to \infty ) \sim 1/r$. The summation over sites of the
simple cubic space-time lattice with periodic boundary conditions,
${\bf r}$, is assumed. Integer currents ${\bf j}_{a{\bf r}} =
(j_{a{\bf r}})_{\mu} $ with $a=1,2$ and $\mu = x, y, \tau $ are
defined on the lattice bonds and are subject to the
zero-divergence, $\nabla {\bf j} = 0$, constraint, i.e. the
configuration space of $j$-currents is that of closed oriented
loops, see Fig.~\ref{fig0}. In terms of particle world lines,
currents in the time direction represent occupation number
fluctuations away from the commensurate filling, and currents in
the spatial directions represent hopping events. In the discussion
of the SF--VBS transition, $j_1$ and $j_2$ currents in the DCP
action represent world lines of spinons which are VBS vortices
carrying fractional particle charge $\pm 1/2$ \cite{dcp1,dcp2}.
The insulating VBS state is characterized by small current loops;
the corresponding spinon order parameter, $\psi_a,\, a=1,2$, is
zero in this phase. When spinon world lines proliferate and grow
macroscopically large, the system enters the superfluid state. The
paired phase (SFS) occurs at large $g$, when only loops in the
${\bf j}_1 - {\bf j}_2$ channel grow macroscopically large. Since
bound pairs of spinons with opposite vorticity make particle and
hole excitations, the SFS state is a superfluid with the solid VBS
order, i.e. a supersolid.

\begin{figure}[tbp]
\begin{center}
\includegraphics[angle=-90, width=0.8\columnwidth]{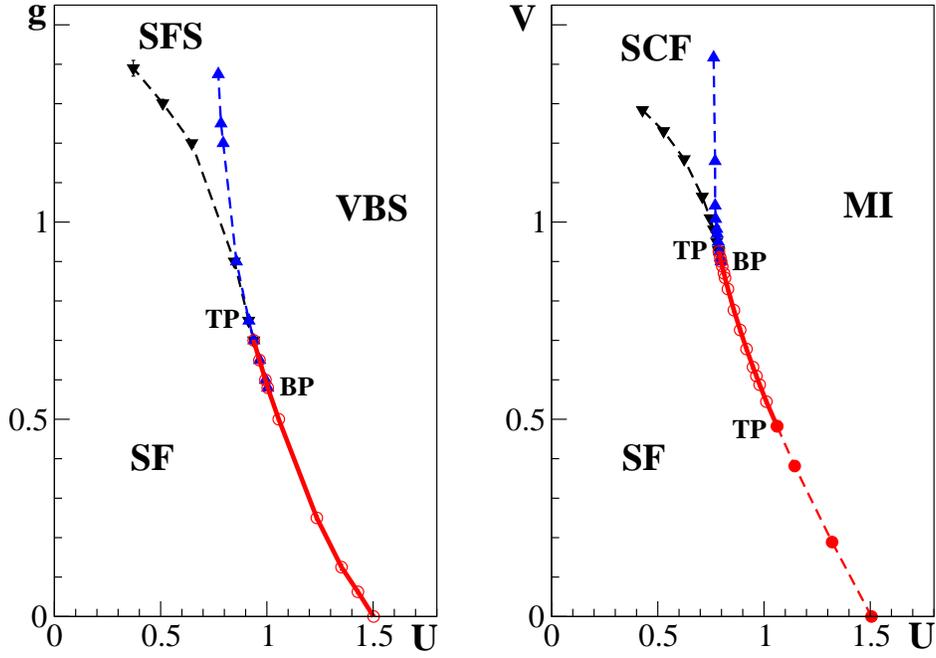}
\caption{(Color online). Phase diagrams of the long-range (left panel),
Eq.~(\ref{DCPA}), and short-range (right panel), Eq.~(\ref{AB}),
actions. Bold solid lines indicate the I-order phase
transition. Error bars are shown but
are typically much smaller than symbol sizes.} \label{fig1}
\end{center}
\vspace*{0.40cm}
\end{figure}

As discussed above, many features of the DCP action are
remarkably similar to those of a more conventional two-component
Bose-Hubbard model \cite{twocolor} with the short-range
interaction potential
\begin{equation}
S_2^{(s)} =U\: {\bf j}_{1{\bf r}}^2 +
U \: {\bf j}_{2{\bf r}}^2 +
V \: \left( {\bf j}_{1{\bf r}}+{\bf j}_{2{\bf r}} \right)^2 \;,
\label{AB}
\end{equation}
with $U>0, \, V>0$.
Phase diagrams for the long- and short-range actions are shown in
Fig.~\ref{fig1}. Different phases are identified in terms of loop
sizes for ${\bf j}_1$ and ${\bf j}_2$ currents similar to the
discussion given above: the Mott insulator (MI) is characterized
by small ${\bf j}_1$- and ${\bf j}_2$-loops, in the two-component
SF there are macroscopically large ${\bf j}_1$- and ${\bf
j}_2$-loops, and in the SCF state only ${\bf j}_1-{\bf j}_2$ loops
are macroscopically large.

\begin{figure} 
\begin{center}
\includegraphics[width=0.8\columnwidth]{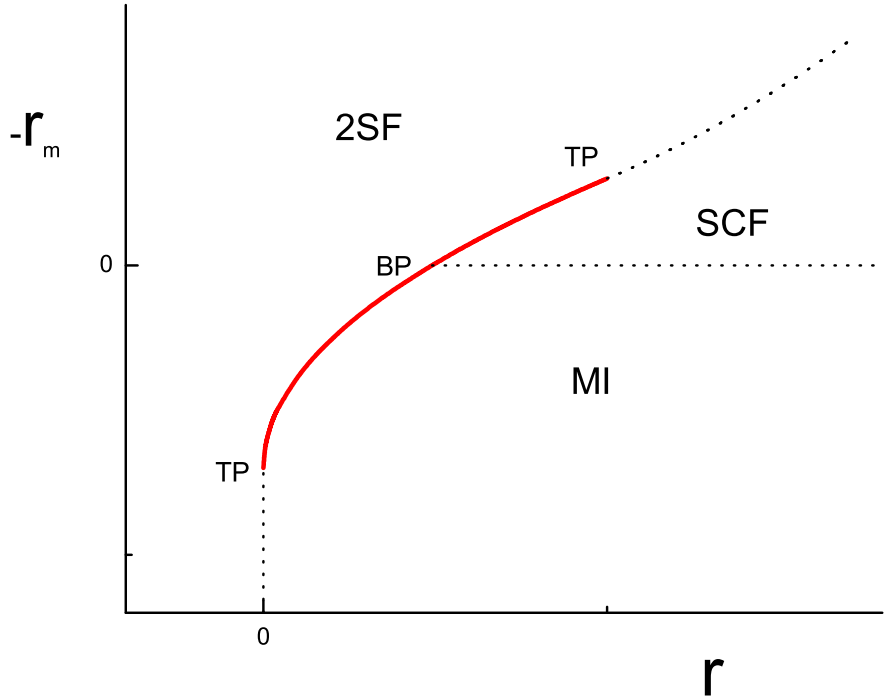}
\vspace*{-0.2cm}
\caption{
(Color online). Sketch of the mean field phase diagram for
the effective action (\ref{Action}). The bold (red) solid line
represents I-order transitions. Dotted lines are used for II-order
transitions. The top of the solid line corresponds to upper TP,
and the bottom -- to lower TP. All three phase boundaries meet at BP.
    } \label{MF}
\end{center}
\vspace*{0.5cm}
\end{figure}

Previous Monte Carlo data for the BH2 model (\ref{AB})
did not agree with the mean field theory
which describes the system in terms of three fields --- $\psi_{1,2}$
for each component and $\Phi$ for the paired field \cite{twocolor}.
Omitting gradient terms the mean-field action is given by:
\begin{eqnarray}
 S_{MF}  =  {1 \over 2}
\big [ r(|\psi_1|^2 + |\psi_2|^2) & + & r_m |\Phi|^2 \big] + {1 \over 4}
\big [U_\psi (|\psi_1|^4
 +   |\psi_2|^4) + U_\phi |\Phi|^4  \big]
\nonumber \\
& - & \gamma (\Phi \psi^*_1 \psi_2 + \rm{c.c.}) \; , \label{Action}
\end{eqnarray}
where $r, \, r_m,\, U_\psi >0, \, U_\phi >0,\, \gamma$ are some
effective coefficients which depend on the bare coupling parameters
$U,\, V$ in the action (\ref{AB}).
Trivial minimization of (\ref{Action}) with respect to the order
parameter fields results in the phase diagram which can be formulated
in terms of two dimensionless variables $\sim r$ and $\sim r_m$, with
all other coefficients rescaled to become unity \cite{twocolor}.
A sketch of the diagram is presented in Fig.~\ref{MF}. It features a continuous
SCF-MI transition, which terminates at BP point, as well as 
I-order MI-SF and SCF-SF transitions.
The SCF-SF transition starts as a first-order line at BP and goes up to
upper TP. Then, it continues as a second (II) order line for larger values of $V$.
Similarly, the MI-SF transition is of I-order type between BP and
lower TP. Then, it continues as a II-order $U(1) \times U(1)$ transition
line towards the $V=0$ point where the components become decoupled.
The length of the I-order line along the SCF-SF boundary is
predicted to be about two times shorter than that along the MI-SF line,
see Fig.~\ref{MF}.

It is worth stressing that the I-order transition is induced by fluctuations of the
paired field $\Phi$. Indeed,
one can approximately integrate $\Phi $ out using gaussian
approximation for this field and generate an effective
interaction term $\sim  - (\gamma^2 /|r_m|) |\psi_1|^2|\psi_2|^2$.
It destabilizes the system for sufficiently small
$|r_m|$ \cite{twocolor}. Thus, the vicinity of BP must feature
a I-order line (along MI-2SF and SCF-2SF). Correspondingly, there are two minimal
values of $(\gamma^2 /|r_m|)_{min}$ -- one for $r_m >0$ and the other for $r_m <0$ -- below which the transition becomes
continuous along the respective boundaries.
It is also important to realize that at the mean-field level there 
is no difference between the
BH2 and DCP models.

Monte Carlo simulations \cite{twocolor} did find an extended I-order MI-SF boundary
and lower TP, but failed to resolve conclusively BP and upper TP.
Standard numerical tools based on studies of hysteresis and
probability distributions for the action were not precise enough for this
system (close to TP the I-order transition is very weak and hard
to identify in finite-size simulations). In all other respects,
the mean field theory and simulations agreed on the topology
of the phase diagram. Another important observation is that simulations
of the same model (\ref{AB}) in $D=3+1$ dimensions \cite{unpubl}
revealed an extended region of first-order transitions above BP.
Since $D=4$ is the upper critical dimension where fluctuations
become essentially insignificant this result was expected.
Thus, an intriguing possibility has emerged that the predicted
first-order transition is destroyed by fluctuations in $D=2+1$.

In this work we apply a new flowgram method to perform a more refined analysis of
the phase diagram in the vicinity of BP point and find that
the SCF-SF boundary does contain a I-order region in $D=2+1$.
The distance between BP and upper TP turns out to be an order of
magnitude smaller than predicted by the mean-field theory. We attribute
this to the fluctuation induced suppression of the pairing effect 
so that higher $V$ is required to produce the SCF phase.

At this point one may ask why not to trust the mean-field theory
on another prediction, namely, the existence of lower TP for
the DCP action. The precise answer lies in how the effective
coefficients in (\ref{Action}) depend on $U,\, g$, and $L$. For
the short-range model, the relation is rather straightforward,
$\gamma \sim \sqrt{V}$, leading to finite value of lower TP at
some $V=V_{LTP}\neq 0$. Ostensibly, the long-range nature of
interaction in (\ref{DCPA}), which culminates in the runaway flow
\cite{Halperin2}, makes the relation between $\gamma$ and $g$
system size dependent through $g_{\rm eff} = gL$ (in leading order
in $g$) and precludes simple perturbative reasoning for lower TP
even for small $g$. In other words, pairing correlations are never
weak at large scales in the DCP action, and it is possible that this
model is always above lower TP of the mean field theory
(\ref{Action}).

\section{Flowgram method}

In this Section we introduce the flowgram method and apply it
to the study of the short-range model (\ref{AB}).
The most fundamental property of current loops responsible for the
superfluid response of the system are winding numbers \cite{Ceperley}
\begin{equation}
{\bf W}_a = L^{-1}\sum_{\bf r} {\bf j}_{a{\bf r}} \;. \label{W}
\end{equation}
Integer $W_{a} $ numbers count how many times ${\bf j}_{a{\bf r}}$
currents wind around the system with periodic boundary conditions.
The superfluid stiffness, $\Lambda_a$, is directly proportional
to $\langle {\bf W}_a^2\rangle /L $. For any scale-invariant transition
there should be $\sim  O(1)$ loops of size $R$ in a volume $R^D$,
which implies that $\langle {\bf W}_a^2\rangle \sim  O(1)$ at
criticality and directly leads to the Josephson relation
$\Lambda_a \sim 1/L$. Moreover, at the continuous transition
point not only the $\langle {\bf W}_a^2\rangle $ value
but also the whole distribution function $F({\bf W})$ is universal
since it is governed by the system behavior at the largest scale.

Given close similarities between the DCP and BH2 phase
diagrams it is  convenient to introduce the following terminology:
the transition line between phases with zero and non-zero ${\bf
W}_{-}={\bf W}_1-{\bf W}_2$ winding number fluctuations is called
a {\it neutral} line (described by proliferation of {\it neutral}
~${\bf j}_{-}={\bf j}_1 - {\bf j}_2$ loops), and the transition
line between phases with zero and non-zero ${\bf W}_{+}={\bf
W}_1+{\bf W}_2$ winding number fluctuations is called a {\it
charged} line (described by proliferation of {\it charged} ~${\bf
j}_{+}={\bf j}_1+{\bf j}_2$ loops in addition to the loops of
${\bf j}_{-}$ which are  already  macroscopic). This terminology
originates from the identification of $j$-currents as world line
trajectories for charged spinon fields in the DCP action; only
${\bf j}_+$-currents are coupled by the long range Coulomb
potential. Using this terminology, the SCF-SF (SFS-SF) transition
is a charged line and the MI-SCF (VBS-SFS) transition is the
neutral line; the two lines merge at BP. Notice that above BP
the neutral winding number fluctuations $\langle {\bf
W}^2_{-}\rangle$ on the charged line must exhibit linear growth
with $L$ because the paired phase is characterized by finite
stiffness (see Fig.~\ref{fig_lin}).
The same language will be used for the short-range action (\ref{AB})
even though no Coulomb potential is involved.

Formally, in Eq.~(\ref{DCPA}) only ${\bf W}_{-}$
numbers may have non-zero values in the thermodynamic limit.
Indeed, charged winding numbers couple to 
$Q_{{\bf q}=0}=1/q^2=\infty$ and thus are forbidden.
However, charged loops of macroscopic size which simply do not wind 
around the system (but are similar to winding loops otherwise)
are allowed and their contribution to the action can be estimated
as $Q_{{\bf q}=1/L}/L \sim L$ (one would get the same result
for loops with non-zero winding numbers if the zero-momentum
compotent of the interaction potential is "regularized":    
$Q_{{\bf q}=0} \to Q_{{\bf q}=2\pi/L}\sim 1/L^2$). 
Since $Q_{{\bf q}=0}=\infty $ imposes a global constraint 
on a single variable of the configuration space of the model 
it may not have any effect on the nature of the scale-invariant 
transition. For example, on approach to the critical point 
charged loops of larger and large size (but smaller then 
$L\to \infty$ ) will proliferate anyway. We remove the global 
constraint by setting $Q_{{\bf q}=0}=Q_{{\bf q}=2\pi/L}$ for only 
one reason---to use non-zero winding number fluctuations as an 
indication of the deconfinement transition. We expect 
$\langle {\bf W}_+^2\rangle \sim 1$ at the transition point. 
The advantage of looking at the ${\bf W}_{+}$ numbers is that they are
sensitive {\it only} to the deconfinement transition to the SF
phase and remain zero in both the VBS and SFS phases.   

The key elements of the flowgram method are:
\begin{itemize}
\item introduce a definition of the critical point for finite-size
systems consistent with the thermodynamic limit and insensitive
to the transition order, \\
\item at the transition point, calculate quantities which are scale-invariant
for the continuous phase transition in question, vanish in one of the phases
and diverge in another; in our case such quantities are statistical averages of
winding numbers squared, $R_{\mp}(L)= \langle {\bf W}_{\mp}^2 \rangle$, \\
\item study the flow of $R_{\mp}(L)$ with system size $L$ for given model parameters.
As far as the thermodynamic critical points are
concerned, the critical values of the  corresponding parameters 
in the action are naturally obtained by extrapolating the 
sequence of finite-size critical points to $L=\infty$.
\end{itemize}
Observing how $R(L)$ scales with $L$ at the transition point
allows studying the order of a transition,  since $R(L\to \infty
)$ has to saturate  at some universal constant for second-order and to
diverge for first-order transitions. Moreover, radical changes in the
flowgram from saturation to divergence with an unstable separatrix
in between as a function of coupling  parameters is a clear sign
of TP. It is important to emphasize that the flowgram method
does not rely on detecting energy barriers between competing
phases. It is simply based on phase coexistence typical for 
first-order transitions.

We use the following definition of a transition point:
for any given set of $(g,L)$ or $(V,L)$ we determine the critical
value of $U$ from the condition
that the ratio of probabilities of having zero and non-zero winding numbers is
some fixed number $C$ of the order of unity:
\begin{equation}
C_+=F(0)/ \sum_{{\bf W}_- \neq 0} F({\bf W}_-)= C \;,\quad C_-=F(0)/ \sum_{{\bf W}_+ \neq 0} F({\bf W}_+)=C,
\label{crit_def}
\end{equation}
for the neutral and the charged lines, respectively.
We use this particular criterion (among many others) because in
the thermodynamic limit it gives exact answer for both
second- and first-order transitions, since the ratio of probabilities
in (\ref{crit_def}) is zero in the superfluid phase and infinity
in the insulating phase \cite{note2}. 
As mentioned above, at the critical point we compute $R_{\pm}$.
For the scale-invariant continuous transition one expects this quantity
to approach its universal value as $L \to \infty$; for the
the I-order transition it has to diverge with $L$ since in the superfluid
phase $\langle {\bf W}_{\pm}^2\rangle \sim L$.
The bicritical point is detected by observing vanishing 
charged winding numbers $R_+$ at the neutral line as $L$ is increased, 
and diverging, $R_{-} \propto L$, neutral winding numbers on the charged 
transition line. 

Close to the weakly I-order transition the ratios $C_\pm$ (\ref{crit_def})
and averages $R_{\pm}$ exhibit strong fluctuations and are hard to compute 
precisely. However, fluctuations in $R_{\pm}$ are {\it strongly correlated} 
with fluctuations in $C_\pm$. This property, known as covariance,
can be used to reduce statistical errors \cite{Sandvik2}. We have employed 
covariance, by plotting $R_{\pm}$ {\it vs} $C_\pm$, to reduce error bars 
by nearly an order of magnitude.

The flowgram method also allows to ``visualize" the
renormalization of coupling parameters. If data obtained for
$(g_1,L_1)$ collapse on top of data for $(g_2,L_2)$ then one can
consider this as an evidence that two systems are already in 
the scaling regime and are governed by the
same effective coupling at large scales, $g_{\rm
eff}(g_1,L_1)=g_{\rm eff}(g_2,L_2)$. In practice, we plot data for
$R_{\mp} $ as a function of $\ln (L/ \xi(g) )$ where $\xi(g)$ is
some interaction-dependent length-scale, i.e. we study if system
properties at large scales are identical for different coupling
constants $g$ up to the length-scale renormalization and are part
of the same renormalization flow. If true, all data points for the
DCP model should collapse on a single master curve. In the limit
of small $g$, the runaway flow starts as $g_{\rm eff}(L) \propto
gL$ \cite{Halperin2}, thus the length scale renormalization should
follow the $\xi(g) \propto g^{-1}$ law.

\subsection{ Flowgrams for the short-range model}
 The most controversial
region on the phase diagram for the BH2 model (\ref{AB}) is located in the vicinity
of BP, see Fig.~\ref{fig1}, where previous simulations failed
to resolve the difference between BP and upper TP.
The standard feature of I-order transitions -- a double-peak
shape of the probability distribution for energy/action  $P(S)$ --
works reliably only if there is a significant barrier between two
phases. On approach to TP the barrier vanishes and the double-peak
distribution method fails to work.

Using flowgrams one can identify the first-order transition even if no
distinct discontinuity in system properties can be clearly seen.
The cornerstone of the first-order transition is phase coexistence. In
our case these phases are the insulator characterized by loops of
small size (in the corresponding channel) and the superfluid in
which the number of proliferated loops grows with the system size,
so that the mean square winding number $\langle W^2\rangle \sim L$
for large enough sizes. The statistical properties of such mixture
are drastically different from those of the second-order transition
point characterized by scaling behavior, where the same number of
loops of size $\sim L$  must be seen regardless of the (large) 
system size. The tricritical point is then characterized by 
the unstable separatrix separating the two limiting behaviors.

\begin{figure} 
\begin{center}
\includegraphics[width=0.8\columnwidth]{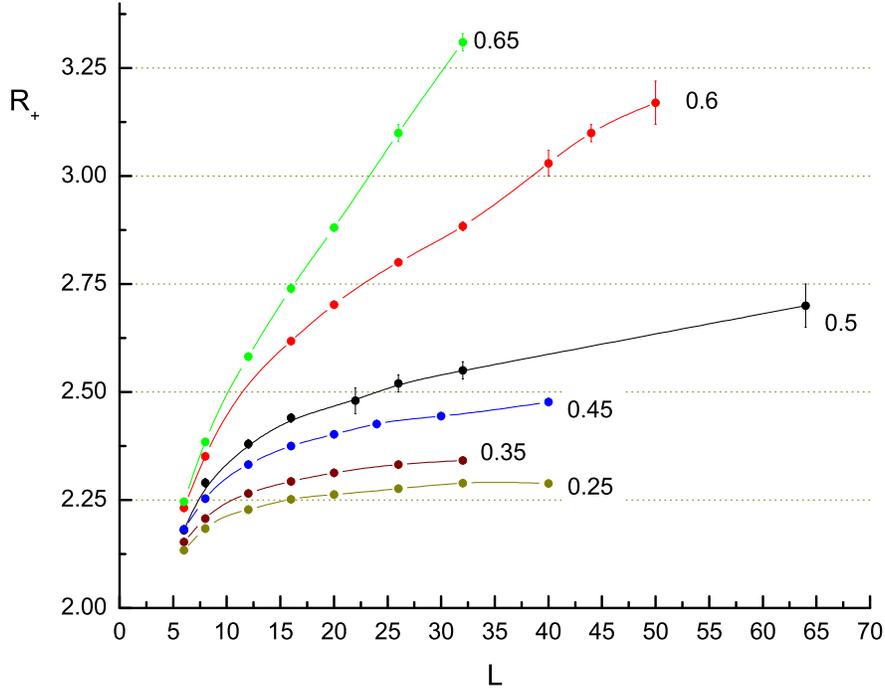}
\vspace*{-0.3cm}
\caption{(Color online). Flowgrams for $R_+$ along the MI-SF line
of the short-range model. The separatrix is located at lower TP
given by $V=V_{LTP}\approx 0.5$.
    } \label{figg1}
\end{center}
\end{figure}

\begin{figure} 
\begin{center}
\includegraphics[width=0.8\columnwidth]{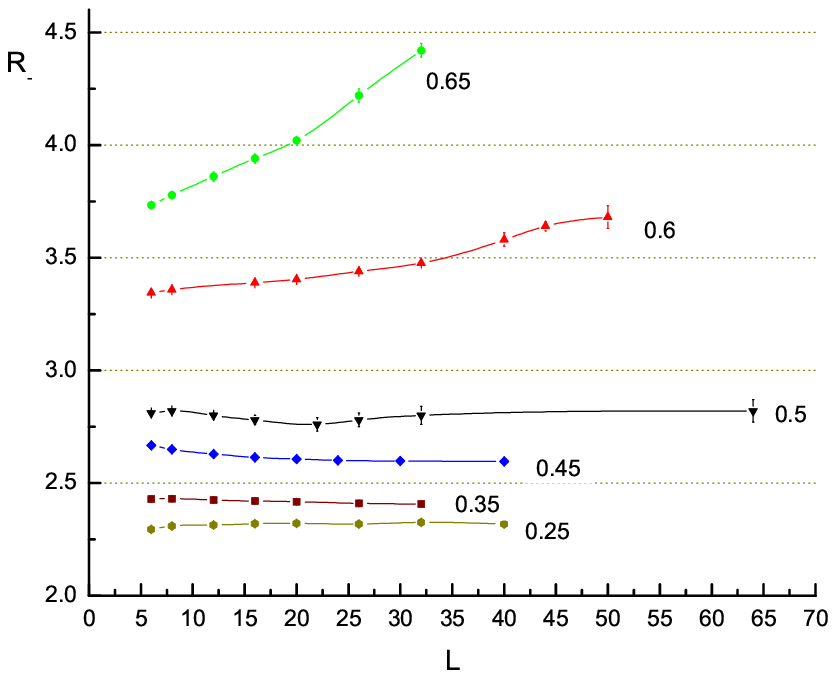}
\vspace*{-0.3cm}
\caption{(Color online). Flowgrams for $R_-$ along the MI-SF line
of the short-range model. The separatrix is located at lower TP
given by $V=V_{LTP}\approx 0.5$.
    } \label{figg1A}
\end{center}
\vspace*{0.50cm}
\end{figure}

In Fig.~\ref{figg1}, we show flows of $R_+$ vs $L$ for different
values of $V$ along the MI-SF boundary. The separatrix position at
$V=V_{LTP}\approx 0.5$ is consistent with lower TP found
previously \cite{twocolor}. In Fig.~\ref{figg1A}, the same separatrix 
is shown to be more pronounced in the neutral channel $R_-$. 
Below we will contrast these flows with the
corresponding plots for the DCP action.

\begin{figure} 
\begin{center}
\includegraphics[width=0.8\columnwidth]{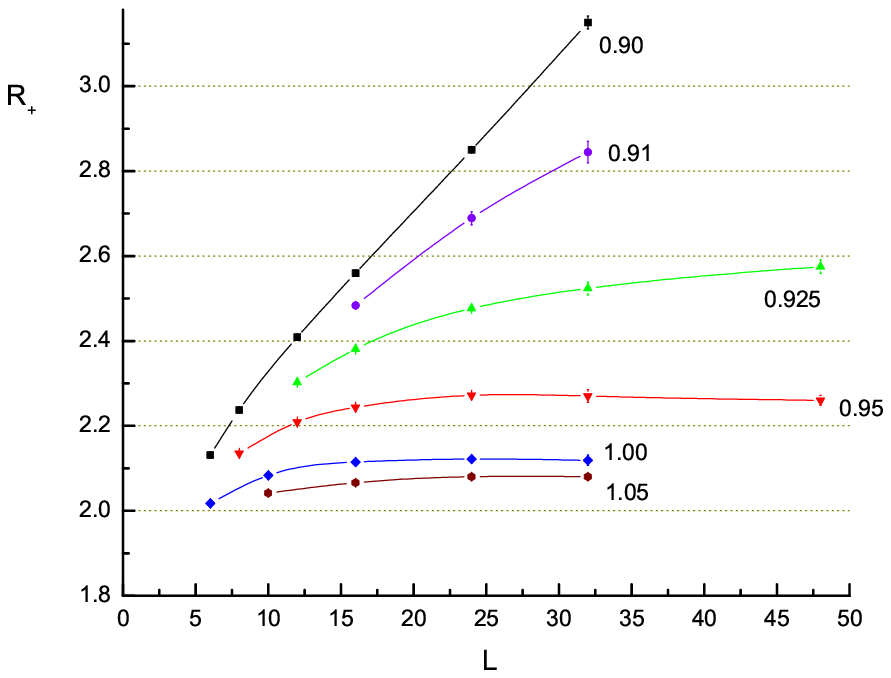}
\vspace*{-0.3cm}
\caption{(Color online). Flowgrams for  $R_+$ along the charged
SCF-SF line of the short-range model. The upper tricritical point
is identified from this plot to be at $V_{UTP}\approx 0.925$. }
\label{figg2}
\end{center}
\end{figure}

Linear scaling $R_{+} \sim L$ is quite obvious close to upper TP as well,
where no distinct double-peak probability distributions $P(S)$ were
seen, see Fig.~\ref{figg2}.
As the interaction strength $V$ crosses the upper tricritical
point located at $V_{UTP}\approx 0.925 \pm 0.01$ the behavior
changes from $R_+ \sim L$ to   $R_+ \to const$.
\begin{figure}[tbp]
\begin{center}
\includegraphics[width=0.8\columnwidth]{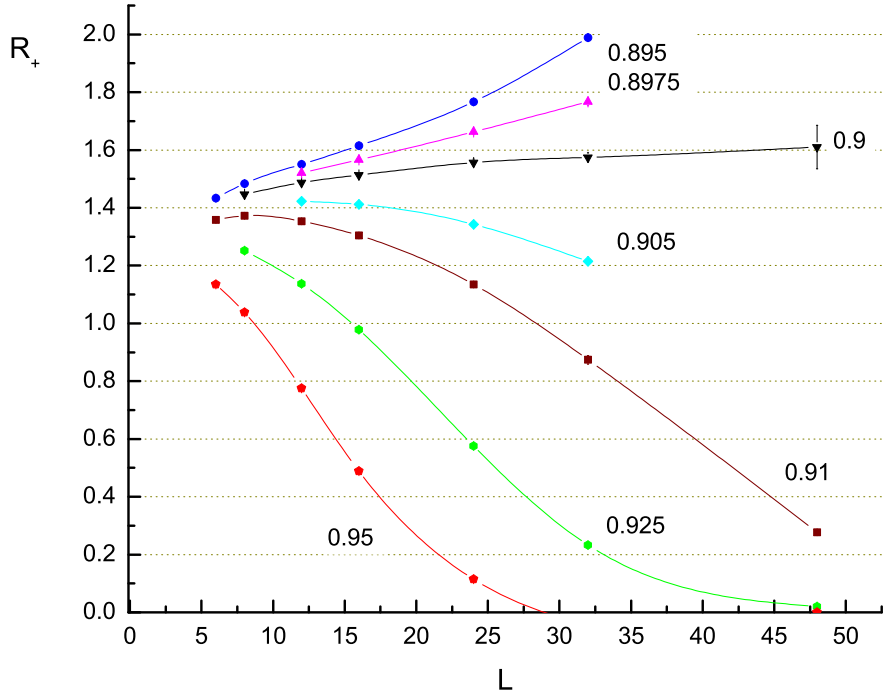}
\vspace*{-0.3cm}
\caption{(Color online). Flowgrams for $R_+$ along the neutral MI-SCF line of
the short-range model. The bicritical point is identified from this plot
to be at $V_{BP}=0.90$.
} \label{figg3}
\end{center}
\end{figure}
\begin{figure}[tbp]
\begin{center}
\includegraphics[width=0.8\columnwidth]{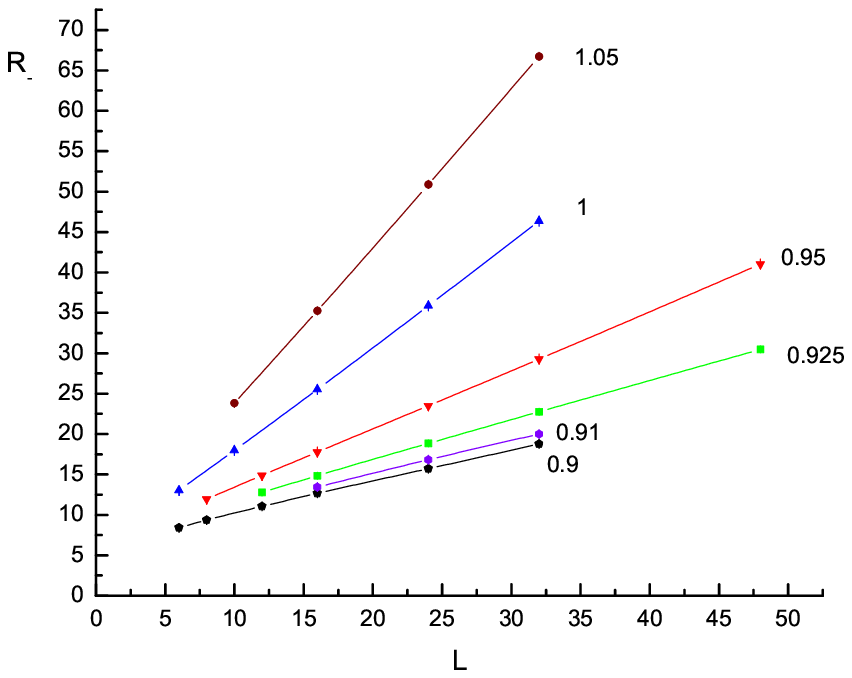}
\caption{(Color online). Flowgrams for $R_-$ along the charged MI-SF line of
the short-range model in the SCF phase above the bicritical point ($V_{BP}=0.90$).
} \label{fig_lin}
\end{center}
\end{figure}

For the determination of BP, in Fig.~\ref{figg3} we
present the flowgram of charged winding numbers on the neutral
transition line close to the point where the MI-SF transition splits
into two. For $V_{LTP}<V< V_{BP}=0.90 \pm 0.005$ we observe
diverging behavior typical for I-order transitions.
For $ V> V_{BP}$ the flow is towards zero values
as $L$ increases and is indicative of the MI-SCF transition
where only winding numbers in the neutral, or paired,
channel proliferate.
In contrast, the neutral windings $R_-$ diverge as $\sim L$
on the charged transition line (SCF-SF) because above the BP the SCF
superfluid stiffness is finite in the SCF phase. This behavior
is seen in Fig.~\ref{fig_lin}.

\begin{figure}[tbp]
\begin{center}
\includegraphics[width=0.8\columnwidth]{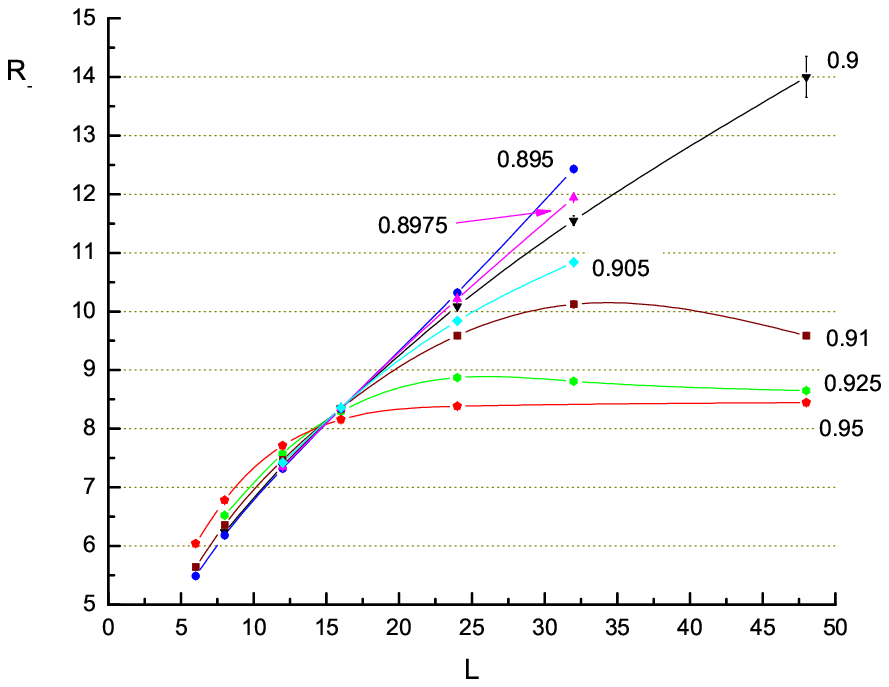}
\vspace*{-0.3cm}
\caption{(Color online). Flowgrams for  $R_{-}$ along the neutral MI-SCF line
of the short-range model. This plot is consistent with the
existence of BP at $V_{BP}=0.90$ (the same value was
deduced from $R_+$ flowgrams) and I-order transitions at
and slightly below $V_{BP}$.}
\label{figg4}
\end{center}
\end{figure}

Fig.~\ref{figg4} shows neutral winding numbers $R_-$ along
the neutral line of MI-SCF transitions in the vicinity
of BP. The crossover from linear scaling $R_- \propto L$
to saturation $R_- \to const$ is very sharp and is
associated with passing through BP, where the II-order
MI-SCF line terminates at the I-order MI-SF/SCF-SF line.
From this plot we estimate $V_{BP}=0.90 \pm 0.005$, in perfect
agreement with the value deduced from flowgrams for the
charged line, see Fig.~\ref{figg3}.

Comparing  $V_{BP}=0.90$ to $V_{UTP}=0.925$ we conclude that
upper TP and BP do not coincide, and the actual phase diagram 
(see Fig.~\ref{fig1inset})
is {\it topologically} identical to the mean field result
\cite{twocolor}. There is, however, a strong (about an
order of magnitude) suppression of the I-order region above the BP
due to strong fluctuation-induced corrections.

\section{Deconfined Criticality}

The field theory of DCP, built on fractionalized spinons (or
vortices in the VBS phase), hints at the remarkable possibility of
a {\it generic} continuous SF-VBS transition. Obviously, 
such II-order transition which occurs between phases characterized by
different broken symmetries cannot be derived from the
conventional Landau expansion in powers of the corresponding order
parameter fields. One cannot even justify the perturbative
expansion because it should start from the fully disordered
quantum groundstate which has no broken symmetries (and
topological orders). There are strong arguments that
disordered quantum groundstate simply does not exist
\cite{dcp3,weak1}. Another option is using a phenomenological
Landau theory which combines superfluid and solid orders into one
multi-component order parameter with non-zero modulus as a
(non-perturbative) constraint \cite{YKIS,weak1}. This approach,
however, predicts that SF-solid transitions are generically of
first-order with the exception of special high-symmetry 
points.

At first glance, DCP predictions are in sharp contradiction with
the Landau approach. However, there is a fundamental difficulty in
the DCP theory, namely, the runaway renormalization flow to strong
coupling which makes the idea of generic continuous SF-VBS
transition speculative and without reliable analytical support
\cite{dcp3}. In the absence of stable perturbative fixed points
other possibilities have to be explored as well. One alternative
is that the DCP action is, in fact, a theory of generic I-order
SF-VBS phase transitions (!) contrary to the original predictions.
Early work on 3D scalar quantum electrodynamics \cite{Halperin2}
did suggest that the renormalization flow leads to the I-order
transition if the number of the matter-field components is
relatively small ($\leq 100$). This prediction, however, is not
taken seriously because it is known to be incorrect for the
one-component system which, by the duality mapping
\cite{Halperin}, can be shown to have a continuous 3D XY
transition. What happens in the multi-component case is a vast
open problem in critical phenomena. It is worth noting, however,
one aspect which is absent in the one-component case and becomes
crucial in the case of two and more components ---it is the
presence of paired phases that may intervene between insulator
and SF. Pairing effect was shown above to induce I-order
superfluid-insulator transition for moderate coupling between
components. Paring correlations in combination with the runaway
flow to strong coupling \cite{Halperin2} would provide a natural
explanation for the generic weakly I-order SF-VBS transitions in
the DCP theory, as conjectured in Ref.~\cite{YKIS}.

The phase diagram for the DCP action shown in Fig.~\ref{fig1} is
not specific to the $j-$current representation. A similar phase
diagram is obtained for the XY phase-gauge field formulation of
the theory Ref.~\cite{Motrunich,motrunich2}. More refined and
extensive simulations have shown that early Monte Carlo results
for the easy-plane model mentioned in Ref.~\cite{Motrunich},
section IIIA, did not go to sufficiently large system sizes to
observe deviations from the power-law scaling and signatures of
the weakly I-order transition \cite{motrunich2}. Since for $g \to
0$ the length scale for observing the I-order transition, if any,
is guaranteed to diverge, one has to be extremely careful in
monitoring deviations from the scaling behavior expected for the
continuous scale-invariant transition. In
Refs.~\cite{Babaev,Motrunich} the second-order SF-VBS transition
with the correlation length exponent $\nu \approx 0.6$ was deduced
using straightforward power law fits of the data for system sizes
$L \le 24$. To make a point of comparison, we studied $\nu$ for
the short-range model (BH2) along the SF-MI line for system sizes
$L\le 128$ using finite-size scaling of the superfluid density
derivative, $\Lambda_s'(U=U_c)\propto L^{1/\nu -1}$. We found that (i)
the scaling curves are nearly perfect straight lines on the
log-log plot from the BP all the way to the lower TP, (ii) good
scaling with $\nu \approx 0.4$ was observed up to $L=32$ {\it in
the middle of the I-order line(!)}, and (iii) in the upper half of
the continuous $U(1) \times U(1)$ SF-MI line the slope of the best
linear fit produced a set of interaction dependent $\nu (V)$ in
the range between $0.67$ and $0.5$. This set of data is indicative
of the fact that corrections to scaling are extremely important to
take into account even for continuous transitions. With
corrections to scaling included into the fits the data become
consistent with $\nu\approx 0.67$ though the error bars increase
by an order of magnitude. We thus conclude that for two-component 
models $\nu$ cannot be reliably determined on system sizes 
$L\le 24$. More importantly, one may completely miss the 
I-order transition.

In what follows we discuss results of worm algorithm Monte Carlo
simulations performed for the DCP action in the self-dual
representation \cite{Babaev,Motrunich} (equivalent to the
field theory of two identical scalar fields with Abelian gauge
and U(1)$\times$U(1) symmetry of quartic interactions),
construct its phase diagram, and prove our earlier
assertion that the transition to the SF phase is always either
I-order or through the SFS state \cite{YKIS}.
The supersolid state exists only for sufficiently large values of
coupling to the gauge field $g> g_{BP}$ with $g_{BP} \approx 0.6 $. For
$g<g_{BP}$ we find I-order VBS-SF transition for any finite value of $g$.

\subsection{Bimodal distribution}
\begin{figure}
\begin{center}
\includegraphics[width=0.8\columnwidth]{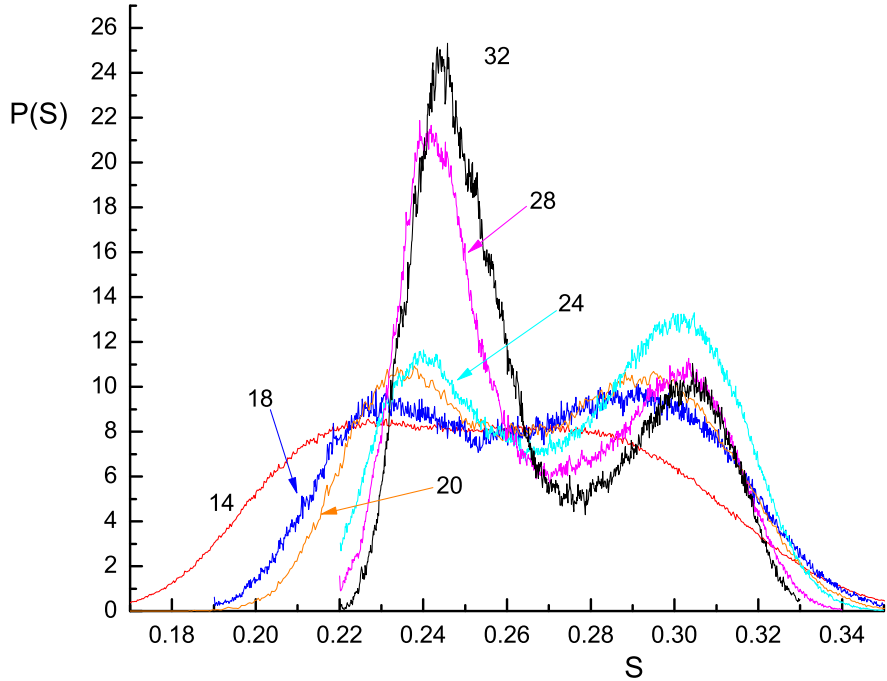}
\caption{(Color online). Normalized probability density
distributions $P(S)$ for the DCP action for $g=0.58$ and different
system sizes; U(L=14)=1.0132, U(L=18)=1.0075, U(L=20)=1.0062,
U(L=24)=1.0041, U(L=28)=1.00329, U(L=32)=1.00263. } \label{fig2}
\end{center}
\end{figure}
\begin{figure}
\begin{center}
\includegraphics[width=0.8\columnwidth]{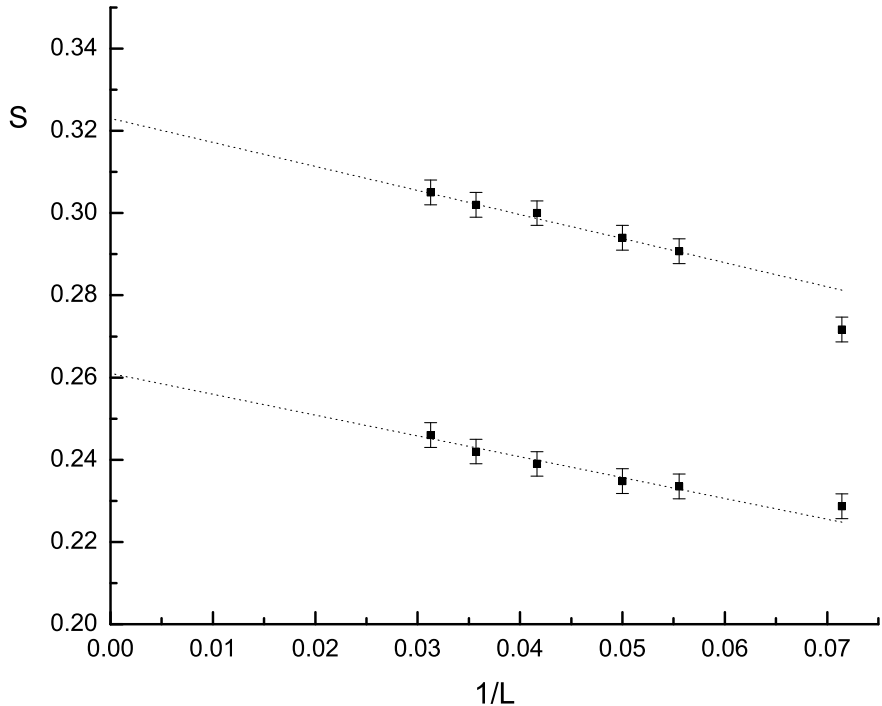}
\caption{ Finite-size scaling of the peak positions
in Fig.~\ref{fig2}. } \label{fig3}
\end{center}
\end{figure}
Before presenting flowgrams for the DCP action and interpreting them
as evidence for the I-order deconfinement transition, we would like
to discuss an unambiguous evidence of the I-order VBS-SF transition
at $g=0.58$. The significance of this point will become clear
shortly after we show that this point can be continuously related to
smaller values of $g$ by the renormalization flow.

In the vicinity of the VBS-SF critical point defined according to
the criterion (\ref{crit_def}) we look at the probability
distribution $P(S)$ and finetune $U$ to the point where this
distribution has the most pronounced double-peak structure or flat
top. Unfortunately, simulations of large $L$ for the long-range
action (\ref{DCPA}) are very time consuming, and $U$ has to be
tuned with accuracy better then $10^{-4}$. Still, we were able to
collect reliable statistics for system sizes $L\le 32$, and have
observed, see Fig.~\ref{fig2}, the development of the double-peak
structure in $P(S)$ starting from an anomalously flat maximum at
$L=14$.

We are not aware of any continuous phase transition which features
more and more pronounced double-peak distribution in energy/action
with increasing the system size \cite{note1}, and consider
Fig.~\ref{fig2} to be an unambiguous evidence in favor of the
I-order transition. The scaling of peak positions with $L$
demonstrating no sign of peaks moving towards each other
is presented in Fig.~\ref{fig3}.

\subsection{Runaway flow and data collapse for small
and intermidiate coupling}

\begin{figure}
\begin{center}
\includegraphics[width=0.8\columnwidth]{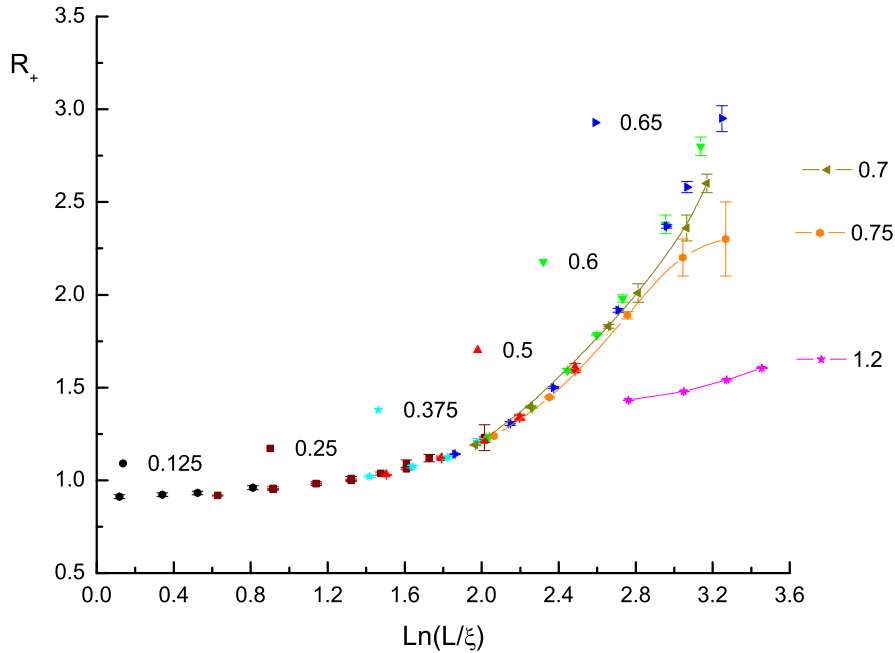}
\caption{(Color online). Rescaled flowgrams for $R_+$ along the
charged line. For $g < g_{BP}\approx 0.58$, this line coincides with
the VBS-SF deconfinement transition, and, for $g > g_{BP}$,
separates SF and SFS phases. System sizes are rescaled by $\xi =
(g+g^2)^{-1}$ so that the data for $g\leq 0.65$ collapse on a single
master curve. The lines connect data points ($g> 0.65 $) which do not collapse.
} \label{fig5}
\end{center}
\vspace*{0.4cm}
\end{figure}

\begin{figure} 
\includegraphics[width=0.8\columnwidth]{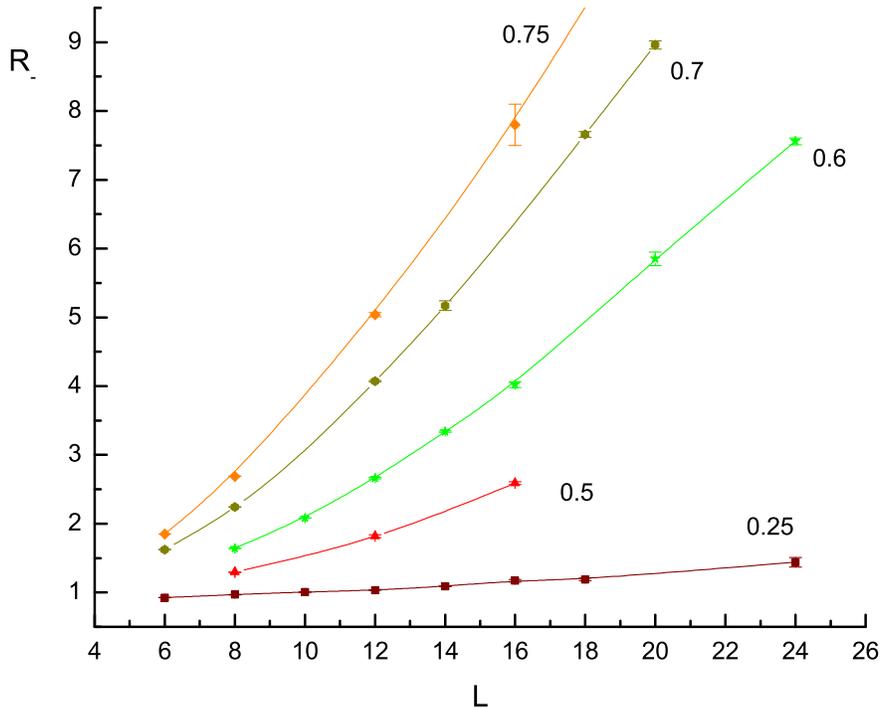}
\vspace*{-0.3cm}
\caption{(Color online) Flowgrams for neutral winding numbers $R_-$
along the charged line demonstrating growth without saturation at
all coupling constants. This behavior is drastically different from
that observed in Fig.~(\ref{figg1A}) for the short-range model, where the
separatrix reveals lower TP.} \label{figg8} \vspace*{0.3cm}
\end{figure}

\begin{figure}
\includegraphics[width=0.8\columnwidth]{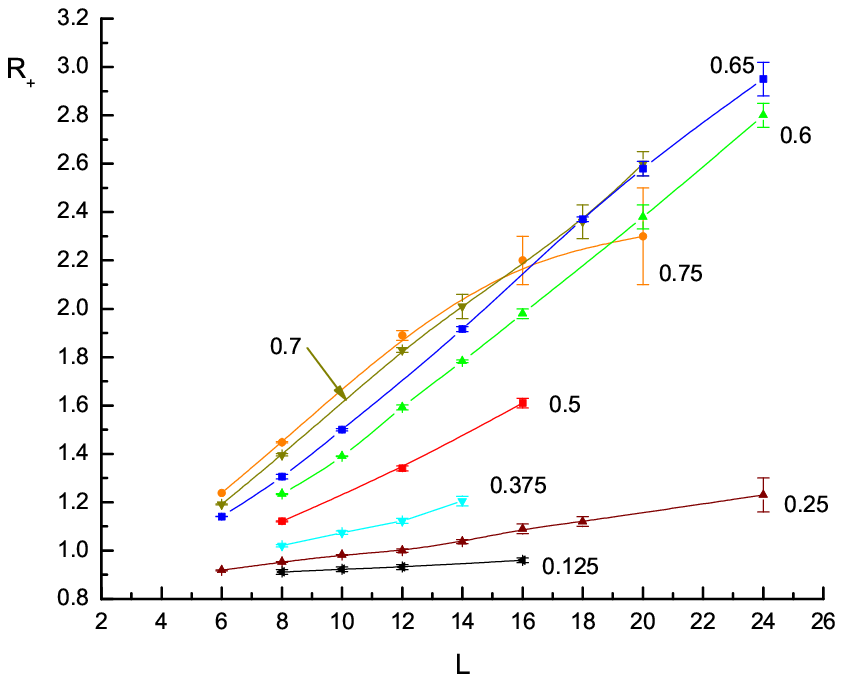}
\vspace*{-0.3cm}
\caption{Flowgrams for charged winding numbers $R_+$ along the
charged line. The same data as in Fig.~\ref{fig5} are now plotted
without rescaling to underline linear divergence of $R_{-}$ with
system size below $g_{UTP} \approx 0.70$. } \label{figg7}
\vspace*{0.4cm}
\end{figure}

In Fig.~\ref{fig5} we present the DCP action (\ref{DCPA}) flowgrams
for various coupling constants
$0.125 \le g \le 1.2$ (the ending point $U=0$, $g\approx 1.6436$ corresponds
to the inverted-XY transition along the charged line)
and system sizes $8 \le L \le 24$.
We observe a remarkable data collapse by rescaling system sizes
using $\xi (g)= (g+g^2)^{-1}$ for values of $g$ up to $g=0.65$
(rescaling is equivalent to the horizontal shift of data points).
The flow starts from the universal value
corresponding to two decoupled $U(1)$ models and, most importantly,
reaches the parameter range ($g=0.58$, $L\ge 14$) where we
observe the development of the double-peak structure in $P(S)$.
Several other features of the flow for $g<g_{BP} \approx 0.58$ are indicative of the
I-order transition. We observe that $R_{-}$, Fig.~\ref{figg8},
and $R_+$, Fig.~\ref{figg7}, grow without
saturation as expected for the mixture of superfluid and
insulating phases. It cannot be reconciled with TP
below $g_{BP}$ since  TP implies an unstable separatrix
and precludes data collapse.

As mentioned in the introduction, data collapse proves that
systems with small $g$ and linear size $L$ have the same
superfluid response as systems with larger $g$ and smaller $L$,
i.e. their long-wave properties are governed by the renormalized
coupling $g_{\rm eff}(g,L)$. We explicitly verify the perturbative
renormalization group result $g_{\rm eff}\propto gL$ of
Ref.~\cite{Halperin2} for small $g$, and thus exclude any
possibility for having TP at $g<0.125$. Renormalization procedure
allows us to extend the flowgram to sufficiently large values of
$gL>18$ and to relate systems with arbitrary small $g$ to the
first-order transition because data points for $g=0.58$ are collapsing
on the same continuous curve. [Simulating $gL=18$ for $g=0.125$
would be equivalent to having $L=115$ -- an impossible task for
the long-range model.] Thus, we conclude that the I-order
transition line goes all the way to the special $g=0$ point which
is described by the $U(1)\times U(1)$ universality for two
independent complex scalar fields

\subsection{Flowgrams along the charged line (SFS-SF transition)}
For $g > 0.65$ the flow collapse
starts to break down. This region features BP and upper TP; for
$g>0.8$ the presence of the paired SFS phase can be seen by the
flowgram method even in small samples. The flow collapse on the
charged line fails because the order of the transition changes at
upper TP ( we estimate $g_{UTP}= 0.70 \pm 0.05 $) which
features an unstable separatrix.  For $g>g_{UTP}$ we are dealing
with the continuous SF-SFS transition of the inverted-XY type. The
diverging flow, then, is replaced with saturation clearly seen already for
$g=0.75$.

In Fig.~\ref{figg8} we present data for flowgrams of neutral winding
numbers along the same (charged) line. This time we do not rescale
distances to verify that $R_-$ obeys the linear law $R_- \propto L$
characteristic of finite superfluid response in the SFS and SF
phases, or phase coexistence at the I-order transition. The
divergent flow observed for $R_-$ at all $g$ on the charged line by
itself is suggesting that BP (where the I-order SF-VBS line splits
into SCF-SF and SCF-VBS lines) is below $g_{UTP} \approx 0.70$:
otherwise, we would observe saturation of $R_-$ between the two
milticritical points. Linear scaling of $R_{+}$ for $g<g_{UTP}$ is
presented in Fig.~\ref{figg7}

\subsection{Flowgrams along the neutral line  (VBS-SFS transition)}

\begin{figure}
\begin{center}
\includegraphics[width=0.8\columnwidth]{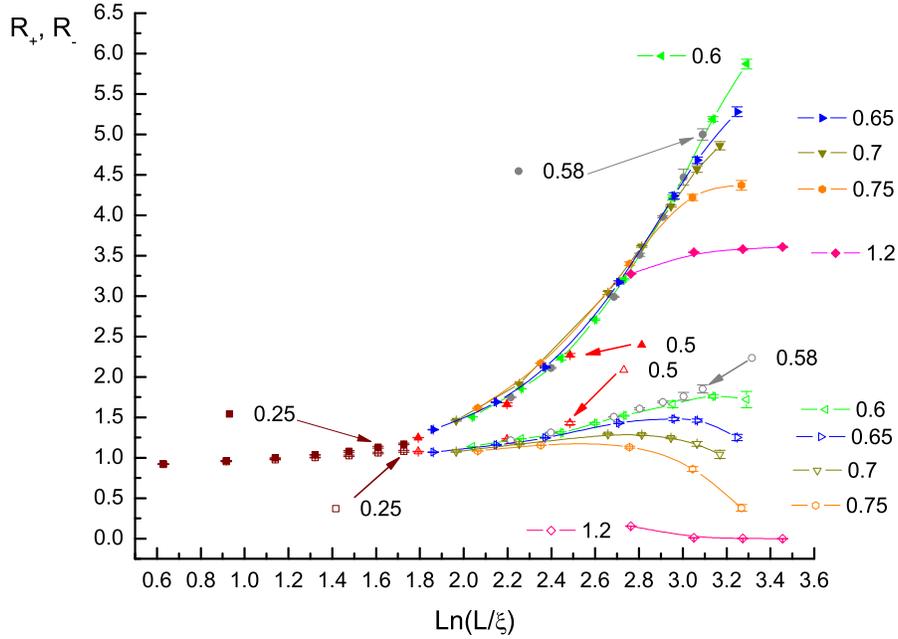}
\caption{Renormalized flowgrams of $R_{\pm}$ along the neutral line.
For $g < g_{BP}\approx 0.58$ this line coincides with the I-order
VBS-SF deconfinement transition and for $g > g_{BP}$ it separates
SFS and VBS phases. Each symbol corresponds to a value of $g$ shown.
The upper set of data points is for neutral
winding numbers $R_{-}$ (solid symbols), and the lower set is for charged ones
$R_{+}$ (open symbols). The lines connect data points ($g \geq 0.58$) 
which do not collapse on the master curve. } \label{figg10}
\end{center}
\vspace*{0.5cm}
\end{figure}
\begin{figure}
\begin{center}
\includegraphics[width=0.8\columnwidth]{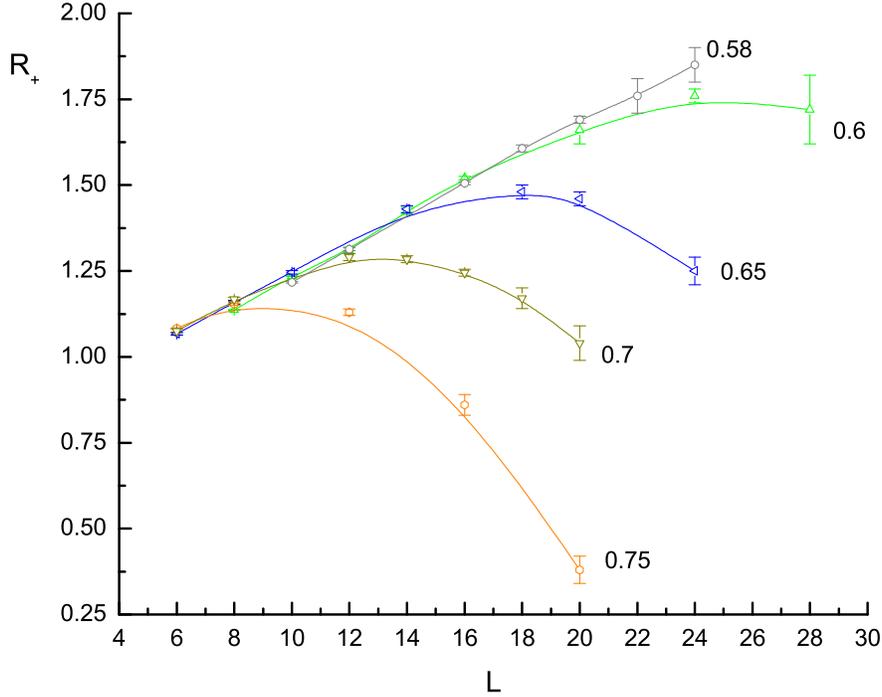}
\caption{Flowgrams for charged winding numbers  $R_+$ along the
neutral line. 
    } \label{figg5}
\end{center}
\end{figure}
\begin{figure} 
\includegraphics[width=0.8\columnwidth]{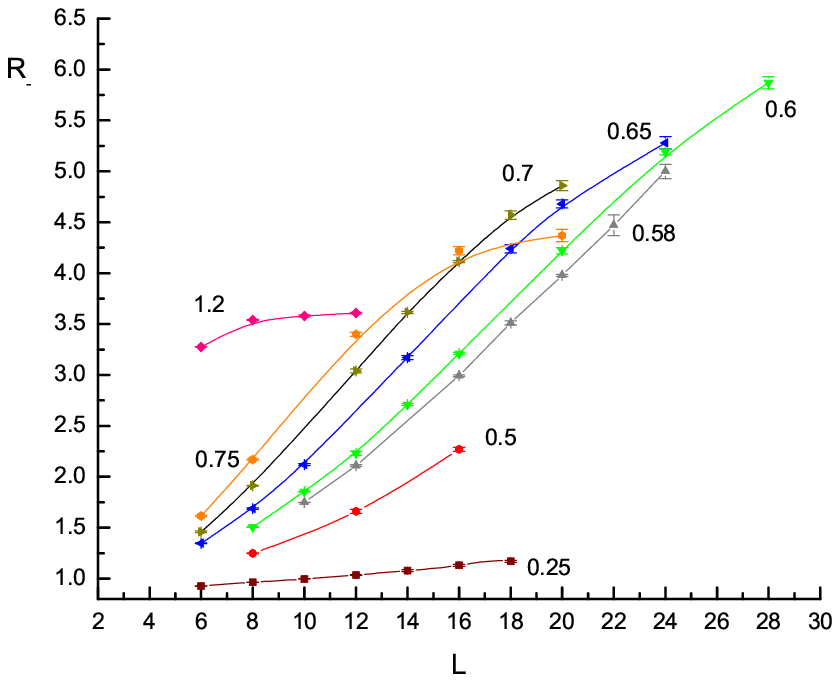}
\caption{Flowgrams for neutral winding numbers  $R_-$ along the
neutral line for $g>g_{BP}\approx 0.58$ and along the VBS-SF
transition for $g<g_{BP}$.The latter set of the points
exhibits linear growth due to the I-order transition as in
Fig.~(\ref{figg8}).
    } \label{figg6}
\end{figure}
Though flowgrams for the charged transition line are sufficient to
determine the topology of the phase diagram, they are rather
insensitive to the location of BP. Instead, one has to look at
flowgrams for the neutral line to observe how it branches out from
the I-order SF-VBS line. These are shown in Fig.~\ref{figg10}. A
nearly perfect data collapse on divergent flow is observed for
both types of winding numbers up to $g\approx g_{BP} \approx 0.58
\pm 0.02$. In Figs.~\ref{figg5} and \ref{figg6} the same data are
plotted without rescaling to underline linear divergence of
$R_{\pm}$ with system size below BP. For higher values of $g$,
winding numbers in the charged channel start decaying to zero and
the flow cannot be collapsed any more. This is an unambiguous sign
that we have passed BP and the neutral SFS-VBS line has branched
out from the I-order VBS-SF transition. At the same time, though
less dramatically, the neutral winding numbers turn to saturation,
as expected for the continuous XY transition.

\begin{figure}
\begin{center}
\includegraphics[angle=-90, width=0.8\columnwidth]{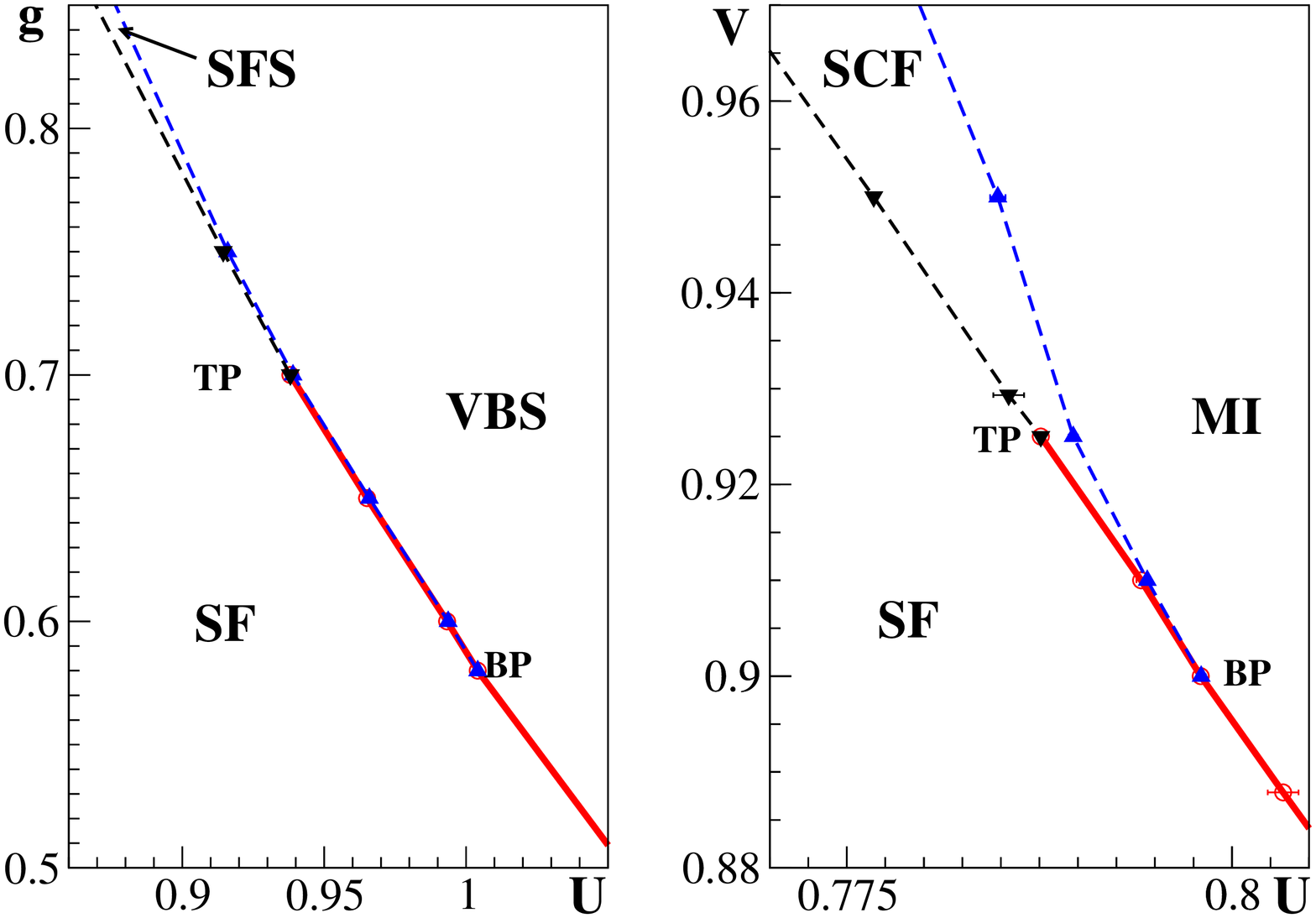}
\caption{(Color online). Fine structure of phase diagrams for the
long-range (left panel), Eq.~(\ref{DCPA}), and short-range (right
panel), Eq.~(\ref{AB}), actions in  the vicinity of BP. Bold solid
lines indicate the I-order phase transition. } \label{fig1inset}
\end{center}
\vspace*{0.4cm}
\end{figure}
We summarize our results for the phase diagrams of the DCP and BH2
actions in Fig.~\ref{fig1inset} which is showing fine details
invisible in Fig.~\ref{fig1}. One may not miss that SF-SFS and
SFS-VBS lines go extremely close to each other between upper 
TP and BP. It is only due to the flowgram method that we were 
able to resolve them conclusively.
The most important result we learn
from Fig.~\ref{fig1inset} is that pairing fluctuations, at the
qualitative level correctly captured by the mean-field theory,
always place the BP below the upper TP regardless of the interaction range.

\section{Conclusion}
To conclude, we have shown that the runaway renormalization flow
to strong coupling in the U(1)$\times$U(1) symmetric two-component
scalar quantum electrodynamics in $D=2+1$ leads to a I-order
deconfinement transition to the SF state. This result, and the
structure of the phase diagram for relatively strong coupling
between the spinons exclude the possibility of generic continuous
VBS-SF transition in the self-dual deconfined critical action. The
I-order transition is fluctuation induced---it develops at
progressively larger length scales as $g \to 0$.

Though in this work we considered a particular DCP model, Eq.~(\ref{DCPA}),
our results have a certain degree of universality since the I-order
transition was found to be very weak and fully developed only in
large systems with more than $10^4$ sites (all results for small $g$ are
universal since they are subject to the renormalization flow).
Thus, our study does rule out the possibility of
having continuous VBS-SF transitions in models with weak-to-intermediate
long-range coupling. Yet one may argue that our simulations do 
not rule the possibility of continuous VBS-SF transition 
in some part of the phase diagram of some yet unexplored model, 
say, for large couplings. Of course, they do not. 
But there is no evidence or even good theoretical argument 
that such a transition line will not be eliminated by the tendency 
to pairing at large $g$. Such model, if any,
would rather be an exception; the generic situation is 
that the combination of the runaway flow  and pairing 
correlations drives a system into either the first-order 
transition for weak and intermediate couplings
or to the paired phase for large ones.
At the phenomenological level, we think that the
strongest argument against DCP is Landau theory based on the
multi-component order parameter with non-zero modulus---so far, all
numerical data for various microscopic models of the VBS-SF 
transition are consistent with it.

We are grateful to E. Vicari, O. Motrunich, S. Sachdev, M. Fisher,
T. Senthil, A. Vishvanath, Z. Te\v{s}anovi\'{c} for valuable and
stimulating discussions. The research was supported by the National
Science Foundation under Grants No. PHY-0426881 and PHY-0426814,
the PSC CUNY Grant No. 665560036, and by the Swiss National Science Foundation.
The CUNY Supercomputer Grid, in general, and the 
CSI Linux Supercomputer Cluster, in particular, played an important role in the project.
We also acknowledge hospitality
of the Aspen Center for Physics during the Summer 2005 Workshop on
Ultracold Atoms.


\begin{thebibliography}{99}
\bibitem{note2}
Obviously, a choice of $C$ determines
only how fast the thermodynamic limit is reached. For the DCP and the BH2 models we used $C=1$
and $C=4$, respectively.
These values are close to $C\approx 2.2$ at the critical point of
the XY-model.
We verified that changing $C$ within the range from unity to about 10
does not change the qualitative features of the flowgrams.
\bibitem{note1} In certain models of continuous phase transitions
the double peak distribution is observed for small system
sizes but it evolves into the single-peak distribution for large
$L$.
\end{thebibliography}
\end{document}